\documentclass{revtex4}

\usepackage{graphicx} 
\usepackage{amssymb}
\usepackage{amsmath}
\usepackage{color}
\usepackage[dvipsnames]{xcolor}

\begin{document}

\title{Theoretical Analysis of Antineutron-Nucleus Data needed for Antineutron Mirrors in Neutron-Antineutron Oscillation Experiments}
\author{K.V.\,Protasov$^1$, V.\,Gudkov\,$^2$, E. A. Kupriyanova$^3$, V.V.\,Nesvizhevsky$^4$, W.M.\,Snow$^5$ and A.Yu.\,Voronin$^3$ \\
\medskip
{$^1$\,Laboratoire de Physique Subatomique et de Cosmologie, UGA-CNRS/IN2P3, Grenoble, France-38026}\\
{$^2$\,Department of Physics and Astronomy, University of South Carolina, South Carolina, USA-29208}\\
{$^3$\,P.N. Lebedev Physical Institute, 53 Leninsky prospect, Moscow, Russia-119991}\\
{$^4$\,Institut Max von Laue -- Paul Langevin, 71 avenue des Martyrs, Grenoble, France-38042}\\
{$^5$\,Department of Physics, Indiana University, 727 E. Third St., Bloomington, Indiana, USA-47405}
}

\begin{abstract}
The values of the antineutron-nucleus scattering lengths, and in particular their imaginary parts,  are needed to evaluate the feasibility of using neutron mirrors in laboratory experiments to search for neutron-antineutron oscillations. We analyze existing experimental and theoretical constraints on these values with emphasis on low $A$ nuclei and use the results to suggest materials for the neutron/antineutron guide and to evaluate the systematic uncertainties in estimating the neutron-antineutron oscillation time. As an example we discuss a scenario for a future neutron-antineutron oscillation experiment proposed for the European Spallation Source. We also suggest future experiments which can provide a better determination of the values of antineutron-nuclei scattering lengths.

\end{abstract}

\maketitle

\section{Introduction}
Neutron-antineutron ($n-\bar{n}$) oscillations would violate the conservation law for baryon number $B$ by two units. A discovery of $B$ violation holds dramatic implications for particle physics and cosmology \cite{Stueckelberg38,Sakharov67,Kuzmin70,Georgi74,Zeldovich76,Hooft78,Davidson1979,MohapatraPRL80,Kazarnovskii80,Kuo80,Chang80,MohapatraPLB80,Chetyrkin81,Dolgov81,Cowsik81,Rao82,Misra83,Rao84,Kuzmin85,Fukugita86,Shaposhnikov87,Dolgov92,Huber01,Babu01,Nussinov02,Babu06,Dutta06,Berezhiani06,Bambi07,Babu09,Mohapatra09,Dolgov10,Gu11,Morrissey12,Arnold13,Canetti13,Babu13,Gouvea14,Berezhiani16,Grojean18,Berezhiani2019,Shrock2019,Shrock_2_2019,Shrock_3_2019}. 
Experimental searches for $\Delta B=2$ processes with neutrons can proceed by searching for antineutron appearance in a free neutron beam or by looking for the large energy release from antineutron absorption in the nucleus of an underground detector material \cite{Kuo80,Dover83,Dover89,Gal00,Chung02,Friedman08,Abe15}. The best free neutron oscillation searches have used a slow neutron beam passing through a magnetically-shielded vacuum chamber to a thin annihilation target surrounded by a low-background antineutron annihilation detector. The present upper bound on the oscillation time $\tau_{n\to\bar{n}}$ of free neutrons used an intense cold neutron beam at the Institute Laue-Langevin (ILL) \cite{Baldo94} and saw zero candidate events with zero background in one year of operation. 

Let's briefly review neutron-antineutron oscillations for neutron propagation in the presence of matter and an external magnetic field $\vec{B}$. The mixing matrix for this case can be written as
\begin{eqnarray*}\label{mixMatr}
\cal{M}= \left(
\begin{array}{cc}
m_n -\vec{\mu_n} \cdot \vec{B} +V_n & \varepsilon \\
\varepsilon & m_n +\vec{\mu_n} \cdot \vec{B}+V_{\overline{n}} \\
\end{array}
\right) ,
\end{eqnarray*}
where $\varepsilon$ is the off-diagonal mixing term in the effective Hamiltonian for the $n/\bar{n}$ two-state system, $\mu_n$ is neutron magnetic moment, and $V_n$ and $V_{\overline{n}}$ are Fermi potentials for neutron and antineutron in matter, respectively.
The diagonalization of this matrix gives mass eigenstates related to pure neutron $|n>$ and antineutron $|\overline{n}>$ states
\begin{eqnarray*}\label{eigenst}
\left(
\begin{array}{c}
|n_1 > \\
|n_1 > \\
\end{array}
\right)
= \left(
\begin{array}{cc}
\cos \theta & \sin \theta \\
- \sin \theta & \cos \theta \\
\end{array}
\right)
\left(
\begin{array}{c}
|n> \\
|\overline{n}> \\
\end{array}
\right)
\end{eqnarray*}
with
\begin{eqnarray*}\label{tan}
\tan (2\theta) =\frac{2 \varepsilon}{(2\vec{\mu_n} \cdot \vec{B} -V_n+V_{\overline{n}})}.
\end{eqnarray*}
Neglecting for the moment neutron $\beta$-decay and annihilation of antineutrons in the matter, this leads to the probability to find an antineutron at time $t$ starting from an initial pure neutron state at time $t=0$ as
\begin{eqnarray*}\label{prob}
P_{n \overline{n}}(t)=\sin^2(2\theta )\sin^2(\Delta E t/2),
\end{eqnarray*}
where
\begin{eqnarray*}\label{delE}
\Delta E = \left[ (2\vec{\mu_n} \cdot \vec{B} -V_n+V_{\overline{n}})^2+4(\varepsilon)^2 \right]^{1/2}.
\end{eqnarray*}

All free neutron oscillation searches so far have arranged for the  neutrons to avoid interactions with matter ($V_n$ and $V_{\overline{n}}$ as small as possible) and the external magnetic field $\vec{B}$ to minimize the energy difference $\Delta E$ between the neutron and antineutron states during the observation time $t$ as the oscillation probability is lowered as $\Delta E$ increases. Even the best magnetic shielding in a perfect vacuum leaves a large enough residual magnetic field that the difference in the neutron and antineutron Zeeman energies $\Delta E \gg \varepsilon$, where $\varepsilon$ is the off-diagonal mixing term in the effective Hamiltonian for the $n/\bar{n}$ two-state system. If the ``quasi-free'' condition $(t\Delta E/\hbar)<1$ is met, where $\hbar$ is the reduced Planck constant, the relative phase shift between the $n$ and $\bar{n}$ states, $e^{-i\Delta Et/\hbar}$, is small enough that the oscillation probability still grows quadratically with $t$ for short observation times and therefore the sensitivity of the measurement is preserved. The practical experimental figure of merit for a free neutron $n - \bar n$ search using this approach is $N_n t^2$, where $N_n$ is the total number of free neutrons observed in the experiment and $t$ is the observation time for free neutron propagation under the quasifree condition.
 
We recently proposed a new concept \cite{NGPSV_Conf,NGPSV_PRL,NGPSV_PLB} which can help realize future laboratory experiments to search for neutron-antineutron ($n - \bar n$) oscillations of free neutrons. It relies on the use of reflection of a $n - \bar n$ superposition from the walls of a specially designed neutron and antineutron ($n$ and $\bar n$) guide. If the  $n$ beam is prepared so that the maximum transverse $v_\perp$ velocities are small compared to the critical velocities $v_{\mbox{\tiny crit}}$ of the guide material for both $n$ and $\bar n$, we showed that in this limit the oscillations are not suppressed and the development of the oscillations remains in the quasifree regime needed for high sensitivity in the measurement. Such coherent reflections of the $n - \bar n$ superposition can allow one to increase the length of the flight path and therefore the observation time compared to a $n/\bar n$ experiment without a guide. In the quasifree regime the oscillation probability increases as the square of the oscillation time. Furthermore with the use of such a guide the lateral size of the installation can be reduced along with the associated cost and difficulty of implementation. Whether or not this option can improve the experimental figure of merit compared with designs similar to that used in the most sensitive neutron-antineutron oscillation experiment conducted at the ILL, which did not use guides in the quasifree propagation region of the apparatus, depends on the detailed neutron source energy and momentum spectrum and experimental setup. The relation of the phase space distribution of the neutrons emitted by the source to the phase space acceptance of the experiment as defined in part by the neutron guides is especially important.  In our previous work we developed the formalism needed for the experiment design and analysis of the results of such an experimental approach.

To optimize the design of such an experiment and to correctly extract the probability of $n - \bar n$ oscillations, it is of fundamental importance to know the values of the scattering lengths $b_{\bar n A}$ of the $\bar n$ on the nuclei of the material of the walls of the $n/\bar n$ guide walls. As there are no direct experimental measurements of $b_{\bar n A}$, these values must be obtained from theory.  An uncertainty in evaluating antineutron-nuclear scattering lengths contributes a systematic uncertainty in estimating the sensitivity of such an experiment and also can preclude the full optimization of the design of the experiment. 

If the observation time $\tau_{\mbox{\tiny obs}}$ of the $n/\bar n$ superposition in a horizontal cold neutron guide is noticeably shorter than its lifetime in the guide, the impact of the present significant uncertainty in the knowledge of $b_{\bar n A}$ can be quite limited as we will show below by explicit examples. The lifetime is determined by the annihilation of the $\bar n$ in the walls of the $n/\bar n$ guide and the phase shift between the $n$ and the $\bar n$ when they are reflected from the walls of the $n/\bar n$ guide. Since the characteristic lifetime due to annihilation of $\bar n$ inside the $n/\bar n$ guide is $\tau_{\bar n} \sim $2 s \cite{NGPSV_PRL}, the observation time should be a few times shorter than this value. This condition can be met for the case of a new experiment now under discussion to search for $n -\bar n$ oscillations at ESS, Lund, Sweden \cite{Addazi2020}. With a neutron guide length of $L_{\mbox{\tiny guide}} \sim 300$ m and a mean neutron velocity of $v_n \sim 750$ m/s, the mean observation time will be $\tau_{\mbox{\tiny obs}} \sim 0.4$ s. This value is smaller than the characteristic lifetime by a factor of $ \tau_{\bar n}/\tau_{\mbox{\tiny obs}} \sim 5$.

We will show in Section 5 that for a typical uncertainty of the imaginary part of the antineutron scattering length $\Delta \mbox{Im } b_{\bar n A}/ \mbox{Im } b_{\bar n A} = \Delta \mbox{Im } a_{\bar n A}/ \mbox{Im } a_{\bar n A} \sim 0.1$, the uncertainty for estimating $\tau_{n \bar n}$ is $\sim 0.5\%$. As usual, we denote $a$ scattering length on free and $b$ on bound nucleus $a= \frac{A}{A+1}b$, $A$ being the nuclear atomic weight. Even if this knowledge was an order of magnitude worse it would not lead to unacceptable loss of experimental sensitivity. Still the analysis of the knowledge of $b_{\bar n A}$ should be carried out in the most conservative manner possible to completely exclude the possibility of incorrect interpretation of results of such a fundamental experiment. 

Measurements of $b_{\bar n A}$ values are only available for hydrogen and helium. The experimental data on antineutron-nucleus interaction at low energies are scarce and possibly (perhaps probably) contradictory,  and there is no generally accepted theoretical model able to describe the whole set of experimental data. We evaluate the limits on the imaginary part of the antineutron-nucleus scattering lengths ($\mbox{Im } a_{\bar n A}$) from theory in this paper. There are two ways to obtain information about $\mbox{Im } a_{\bar n A}$ values:

\begin{itemize}

\item To extract them directly from experimental data as done for $n A$ interaction. The data on $\bar n A$ interaction is poor and their interpretation is model dependent. Only $\bar n A$ annihilation cross sections for some nuclei have been measured down to energies corresponding to 70 MeV/c of $\bar n$ momentum in the laboratory frame. Thist is not sufficient to make a partial wave analysis and extract the scattering length. There seems to be no possibility of direct experiments in the foreseeable future as one cannot produce extremely low energy $\bar n$ beams.

For very light nuclei ($A=1 – 4$) more experimental data are available on $a_{\bar p A}$ which can be used to calculate $a_{\bar n A}$. One must correct for isospin symmetry violation (the nuclear interaction is the same for $\bar p$ and $ \bar n$ only for symmetric $N=Z$ nuclei) and 
for the Coulomb interaction, which drastically modifies the $p-A$ cross sections behavior at low energies. 

\item To develop a theoretical model which describes ${\bar p A}$ experimental data and switch off the Coulomb interaction to obtain $a_{\bar n A}$. Most of the data consists of antiprotonic atom data and total annihilation cross sections for $10-100$ MeV antiprotons. Unfortunately no existing model describes all experimental data as we will show. 

\end{itemize} 

In this paper we analyze theoretical models which describe the existing experimental data to understand how accurately $\mbox{Im } a_{\bar n A}$ values can be extracted. We do not discuss $\bar p p$ and $\bar n p$  where experimental information and theoretical calculations are more abundant but not suitable for our purposes. We consider deuterium ($^2$H) for which the scattering length can be extracted from experimental data and light nuclei with $A \sim 10 - 20$, in particular C, for which there are some data both on ${\bar p A}$ and ${\bar n A}$ interaction from antiprotonic atom, annihilation cross section, and differential elastic cross section measurements. Heavier nuclei are not discussed in this paper. With heavier nuclei more partial waves are present even near zero energy, and it becomes impossible to separate them. For heavy nuclei it is also necessary to take into account relativistic effects. 

In addition to the main application to antineutron mirror reflection, our results are also of interest for the analysis of coherent neutron and antineutron propagation in gas \cite{Gudkov2020}, which is another recently-discussed option for future neutron-antineutron oscillation experiments. In this case it is coherent forward scattering rather than mirror reflection which is involved, but the same neutron potential and therefore the same scattering lengths are of relevance. 

The article is organized as follows. Section 2 analyzes particular case of $\bar n ^2$H, which is of interest as deuterium is a candidate component material for a $n/\bar n$ guide. Section 3 presents briefly the existing optical model approaches used to describe heavier nuclei and discusses the $\bar n $C system. Section 4 discusses the quality of description of atomic and elastic scattering data by these models. Section 5 gives estimations for future experiments on $n - \bar n$ oscillations. 

\section{Antinucleon-deuteron system}

There is no experimental data for $\bar n ^2\mbox{H}$ system but the $\bar n^2$H scattering length can be estimated using the $\bar p^2$H one
which can be extracted from the experimental data with minimum assumptions. Both atomic shifts and widths 
\cite{Augsburge1999} and annihilation cross section at low energies \cite{Zenoni1999} of the $\bar p^2$H interaction were measured in LEAR experiments with good precision. The $\bar p^2$H scattering length can be extracted \cite{Protasov2000} from these data. The results from atomic measurements
\begin{eqnarray*}
a^{cs}_0(\bar p ^2\mbox{H})= [(0.7\pm 0.2) - i(0.4\pm 0.3)] \mbox{ fm}
\end{eqnarray*}
and annihilation data
\begin{eqnarray*}
\mbox{Im } a^{cs}_0(\bar p ^2\mbox{H})= - i(0.62\pm 0.05) \mbox{ fm}
\end{eqnarray*}
are in agreement. The imaginary part is measured more precisely in the second experiment, so one can give the following estimation for the scattering length
\begin{eqnarray*}
a^{cs}_0(\bar p ^2\mbox{H})=[(0.7\pm 0.2) - i(0.62\pm 0.05)] \mbox{ fm}.
\end{eqnarray*}

The relation between the scattering lengths with $a_0^{cs}$ (corresponding here to $\bar p^2$H system) and without Coulomb $a_0^{s}$ forces (corresponding to $\bar n^2$H) was intensively studied in the 1990s, in particular for
antinuc\-leon-nucleon system. The most general solution was found in a series of works by Popov {\sl et al.} \cite{Mur83,Popov81}:
\begin{eqnarray*}
\frac{B}{a_0^{cs}}= \left(1 - 2b_1 \frac{r_s}{B}\right) \frac{B}{a_0^{s}} - 2\left( \ln \frac{r_c}{B} +c_0 + c_1 \frac{r_c}{B} \right).
\end{eqnarray*}
Here $B=\frac{A+1}{A}\frac{\hbar c}{Ze^2Mc^2}$ is the Bohr radius, $r_s$ is $\bar p p$ effective range, $r_c$ is so-called Coulomb range, $c_0 = 2 \gamma + \ln 2$, $\gamma = 0.5772$ being the Euler constant; and $c_1$ and $b_1$ are some numerical constants depending on the given form of the potential. These parameters can not be extracted from experiment. This is the reason why a simple phenomenological approximation
\begin{eqnarray}
\label{phenomenology_CP92}
\frac{B}{a_0^{cs}}= \frac{B}{a_0^{s}} + C
\end{eqnarray}
with an adjustable constant $C$ was proposed in \cite{CarbProt92} to describe the anitnucleon-nucleon system. 

For antiproton-proton system, this formula works within $1\%$ precision. The precision of this expression decreases as the Bohr radius $B$ increases. Nevertheless, it is expected to work quite well for antinucleon-deuteron system where the Bohr radius is still large enough with respect to all other nuclear length parameters. 

Unfortunately there are no theoretical 
calculations in the literature which allow a comparison of $\bar p^2$H and $\bar n^2$H scattering lengths. To estimate the precision of this formula for the deuteron, we made a simple calculation of these scattering lengths in a complex Woods-Saxon potential with the same geometry ($R_U=2$ fm, $R_W = 0,5$ fm, $a_U = a_W = 0,2$ fm inspired by Kohno-Wise potential \cite{KW}) but corresponding to two different physical situations. We first choose the parameters of the Woods-Saxon potential to reproduce the case of strong annihilation: $U = 20$ MeV, $W= 3000$ MeV, to give a scattering length close to that observed in the atomic experiment
\begin{eqnarray*}
a_0^{cs}= 0.67 - i 0.62 \mbox{ fm}.
\end{eqnarray*}
The corresponding scattering length without Coulomb forces is very similar
\begin{eqnarray*}
a_0^{s}= 0.72 - i 0.78\mbox{ fm},
\end{eqnarray*}
which implies a precision of the formula (\ref{phenomenology_CP92}) of 1.5$\%$. Next we choose the parameters of the potential ($U = 138$ MeV, $W= 60$ MeV) corresponding to a very light annihilation and for which the interaction between atomic levels and nuclear levels is very strong \cite{Popov81}. It is known that in this case, higher order corrections are very important. One obtains
\begin{eqnarray*}
a_0^{cs}= 0.61 - i 0.69 \mbox{ fm}, \hspace{1cm}
a_0^{s}= 0.70 - i 0.79\mbox{ fm}.
\end{eqnarray*}
which implies a precision of the formula (\ref{phenomenology_CP92}) of $\sim 10\%$. Thus we justify that this formula works with a few $\%$ precision, which is comparable to that of experimental data.

By using the same constant
$C = 7.17 + i 1.46$ as in the $\bar p p$ system one obtains
\begin{eqnarray*}
a^{s}_0(\bar n ^2\mbox{H})= [(0.7\pm 0.2) - i(0.79\pm 0.07)] \mbox{ fm},
\end{eqnarray*}
which is a promising value for antineutron guides made from deuterated materials.

This is the spin averaged amplitude. The experimental data do not allow to extract separately doublet and quartet scattering lengths and the existing theoretical estimations \cite{WGN1985,LT1990,LRW1991,Yan2008} do not provide any important difference between these two amplitudes.

Unfortunately, this approach does not work for heavier nuclei due to the presence of higher partial waves near threshold, which make the $S$-wave scattering length inaccessible from experimental data. 

\section{Antinucleon-nucleus optical potential approach}

To describe the existing data and to calculate the scattering length we use the most popular approach allowing to extrapolate theoretical calculations to zero energy and to calculate scattering length is the optical model. Let us discuss the different optical potential models developed in previous work to describe antinucleon-nucleus data.

Batty {\sl et al.} conducted a systematic study of $\bar p A$,  kaon, and hyperon interactions in a series of articles (see \cite{Batty89, Batty95, Batty01} and references therein) which was initially developed to describe energy shifts and widths of the levels in exotic atoms due to the strong interaction. They introduced a phenomenological optical potential proportional to nuclear density with a complex-valued fitting parameter describing the strength of the potential and adjusted to describe the shifts and width of heavy antiprotonic atoms. For heavy elements one deals with the levels with non-zero orbital angular momentum. 

For light symmetric nuclei within this approach, the interaction of $\bar p$ with the nucleus is described using the optical potential $V(z)$ from \cite{FGM2005}. We consider the one-dimensional case (dependence on $z$ only):
\begin{equation}
\label{equ1}
V(z)=-\frac{4\pi\hbar^2}{2\mu}\left(1+\frac{\mu}{M}\frac{A-1}{A}\right)b_0(\rho_n(z)+\rho_p(z)),
\end{equation}
where $\mu$ is the reduced mass of the $\bar p A$ system, $M$ is the $\bar p$ mass, $A=Z+N$ is the mass number of the nucleus, $b_0$ is the scattering length parameter equal to $b_0=1.3+1.9 i$ fm according to \cite{FGM2005}. This choice of $b_0$ gives the depth of the imaginary part of the potential for C $\sim$ 200 MeV. For simplicity of analysis we have omitted relativistic effects important for heavy nuclei as well as other corrections like nuclear folded density used by the authors of this model.

Proton and neutron densities in the nucleus are described using a 2-parameter Woods-Saxon shape:
\begin{equation*}
\label{equ2}
\rho_{p,n}(z)=\frac{\rho_{p_0,n_0}}{1+\exp\left(\frac{z-R_{p,n}}{a_{p,n}}\right)},
\end{equation*}
where
\begin{equation*}
4\pi\int_0^{\infty}\rho_{p}(z)z^2dz=Z,\hspace*{1cm}
4\pi\int_0^{\infty}\rho_{n}(z)z^2dz=N.
\end{equation*}
Here $R_{n}=R_{p}=2.0005$ fm and $a_{n}=a_{p}=0.523$ fm according to \cite{Fricke1995}.

To understand the main features of this model, we perform some numerical calculations for $^{12}$C by using the Schr{\"o}dinger equation with potential~(\ref{equ1}) and a point-like Coulomb interaction including calculations both $a_{\bar p A}$ and $a_{\bar n A}$ parameters.

This model has a very large imaginary part of nuclear potential (imaginary part of $b_0$) and produces results which have a quite simple geometrical interpretation. The strong annihilation practically ``kills'' the wave function inside the nucleus and the results are therefore mostly sensitive to the periphery of potential. The real part has a pure $A^{1/3}$-dependence corresponding to a black sphere. We will refer to this approach as a {\sl model with strong annihilation}. For the scattering length without Coulomb interactions, Re $a_0^{\mbox{\tiny s}} =(1.54 \pm 0.03) \cdot A^{0.311 \pm 0.05}$ fm. 
The imaginary part is practically $A$-independent and proportional to the diffuseness of the potential: Im $a_0^{\mbox{\tiny s}} =(1.00 \pm 0.04) $ fm. Such dependence is typical for the case of a potential with very {\sl strong annihilation}~\cite{Friedman1999} and similar results were also  found for other hadronic-nucleus systems.

As shown in \cite{KPV2000}, the imaginary part of the scattering length is related to the complex optical potential with an exponential-like tail $U(r) \sim \exp (-r/a)$ which tends to
\begin{equation}
\label{strongAnn}
\mbox{Im } a^{\mbox{\tiny s}}_0 \rightarrow -a(\pi - \varphi),
\end{equation}
where $\varphi$ is the phase of complex potential. For the potential \cite{FGM2005} with $b_0 = 1.4 + i1.8$ fm, one obtains the ratio of imaginary part to diffuseness to be equal to 2.23 in agreement with numerical calculations:
\begin{equation}
\label{scLength}
a^{\mbox{\tiny s}}_0=3.22-1.16i \mbox{ fm}.
\end{equation}

\subsection{Antinucleon-nucleus annihilation cross section}

The antinucleon-nucleus annihilation experiments for nuclei with $A \geq 12$ corresponding to the antinucleon momenta  $\geq$ 100 MeV/$c$ and kinetic energy 5.3 MeV were performed mainly at LEAR. The review of antinucleon-nucleus interactions can be found in \cite{Bressani2003}. These data were enriched recently by some of new measurements at the Antiproton Decelerator machine at CERN.

\begin{figure}[h!] 
\center
\includegraphics [width=10cm]{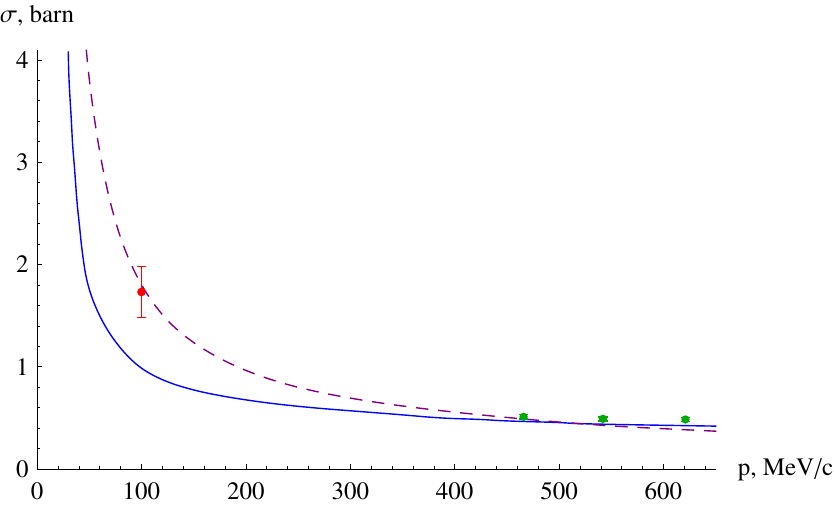} 
\caption{Annihilation cross sections of $\bar p$ on C as a function of momentum (sum of 25 partial waves). The optical potential calculations with usual diffuseness correspond to the solid line; the potential model with double diffuseness calculations are given by the dashed line. Experimental data are used from \cite{Aghai2018} and \cite{Nakamura1984}.} \label{imgPbarC}
\end{figure}

\begin{figure}[h!] 
\center
\includegraphics [width=10cm]{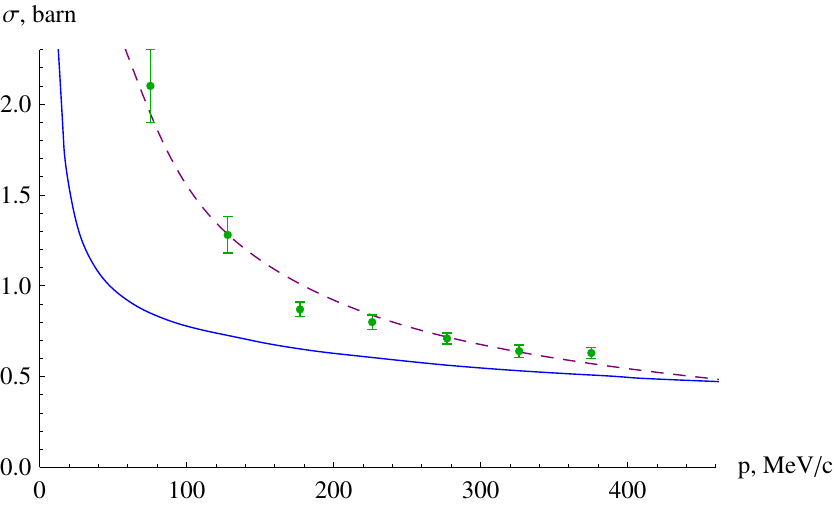}
\caption{$\bar n$C annihilation cross section as a function of momentum. The optical potential calculations with usual diffuseness correspond to the solid line; the potential model with double diffuseness calculations are given by the dashed line. Experimental data are from \cite{Astrua2002}.} \label{imgNbarC}
\end{figure}

In a series of articles Friedman {\sl et al.}~\cite{FGM2005, Friedman2014, Friedman2015} have developed optical models using the discussed above potential, to describe the atomic data. However, the same model can also be used to describe anti\-nucleon-nucleus annihilation and elastic scattering reactions. Unfortunately this potential systematically underestimates the $\bar n A$ annihilation cross section and the $\bar p A$ one at low energies. We present the $\bar p$C annihilation cross section in blue in Fig. \ref{imgPbarC}
and the $\bar n$C annihilation cross section in Fig. \ref{imgNbarC}. 

Since we are interested in very low energies in $\bar p A$  and $\bar n A$ systems, let us recall some important 
modifications in the behavior of nuclear cross sections, which appear in the presence of Coulomb forces,  especially for inelastic (annihilation) cross section
at very low energies. First of all, the famous $1/v$ behavior of the inelastic cross section at low energy is replaced by $1/v^2$. Secondly, all partial annihilations cross sections have the same kinematic behavior and they can be present at very low energies. The scattering lengths without $a^{\mbox{\tiny s}}_0$ and with $a^{\mbox{\tiny cs}}_0$ Coulomb forces are not equal even for the same nuclear potential, as we mentioned in the previous sections.

The partial cross-section of inelastic scattering (annihilation in our case) for neutral particles with angular momentum $l$ is: 
\begin{equation*}
\label{equ4}
\sigma^{(l)}(q)=\frac{\pi}{q^2}(2l+1)(1-|S_l|^2),
\end{equation*}
where $S_l (q)=e^{2i\delta_l(q)}$ for elastic scattering in the presence of inelastic processes is complex with a modulus smaller than one. This simple expression can be generalized for charged particles. We discuss here the case of Coulomb attraction. Following the approach developed in \cite{Mur83,Mur85}, the starting point to obtain the scattering length approximation is the relation between the $K$-matrix for a given orbital momentum $l$ and the strong interaction phase
shift in the presence of Coulomb forces $\delta_l^{cs}$ :
\begin{eqnarray*} 
\frac{1}{K_l^{cs} (q^2) }= g_l(\eta) q^{2l+1} \left[ C_0^2(\eta) \cot \delta_l^{cs} - 2\eta h(\eta)\right],
\end{eqnarray*}
where $q$ is the center-of-mass momentum, $\eta = 1/qB$,
\begin{eqnarray*} 
&&g_0(\eta) = 1, \\
&&g_l(\eta) = \prod_{m=1}^{l} \left( 1 + \frac{\eta^2}{m^2} \right), \hspace{1cm} l=1,2, \ldots \\
&&C_0^2(\eta) = \frac{2\pi \eta}{1- \exp(-2 \pi \eta)}, \\
&& h(\eta) = \frac{1}{2} \left[ \Psi(i\eta) +\Psi(-i\eta) \right] - \frac{1}{2} \ln \eta^2,
\end{eqnarray*}
with the digamma function $\Psi$. 

The $K$-matrix is related to the $S$-matrix by
\begin{eqnarray*} \label{Slsc}
S_l^{cs} (q) = \mbox{e}^{2i\delta_l^{cs}} = \frac{1+ig_l(\eta) q^{2l+1} w (\eta) K_l^{cs} (q^2) }{1-ig_l(\eta) q^{2l+1} w (\eta) K_l^{cs} (q^2) }
\end{eqnarray*}
with $w (\eta) = C_0^2(\eta) + 2i \eta h(\eta)$.

\begin{figure}[h!] 
\center
\includegraphics [width=12cm]{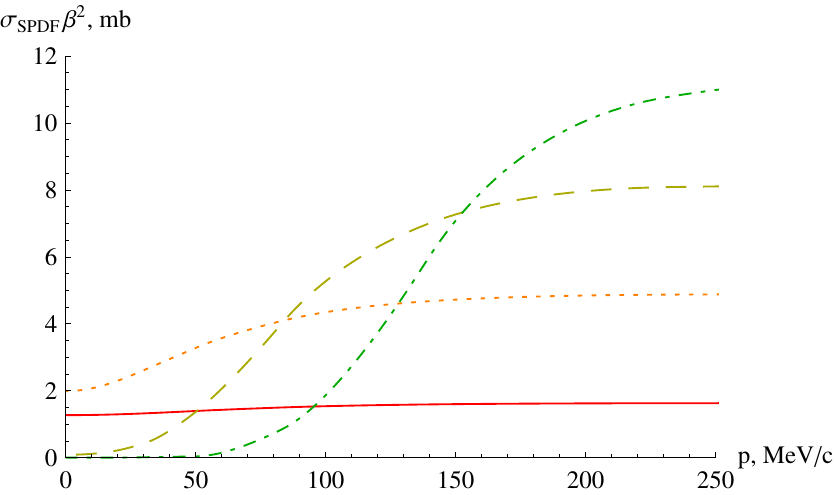}
\caption{S, P, D, F partial cross section $\sigma_l$ multiplied by the square of velocity $\beta^2$ are represented respectively by solid, dotted, dashed, and dash-dotted lines.} \label{img4}
\end{figure}

An example of contributions of different partial waves to the annihilation cross section calculated for the strong annihilation model is shown in Fig. \ref{img4}.
As we can see from this example, P-wave contribution to the annihilation cross section is more important that the S-wave one even at zero
momentum. In this situation, an extraction of S-wave parameters from the experimental data becomes quite a difficult task.

The scattering length approximation used at low energies is equivalent to the replacement of $K$-matrix :
\begin{eqnarray} 
\label{ScatLength}
\frac{1}{K_l^{cs} (q^2) }= - \frac{1}{a_l^{cs}} + \frac{r_l^{cs}}{2} q^2+ o(q^4)
\end{eqnarray}
by a constant, the first term in the expansion. Within this approximation, the partial annihilation cross section $\sigma^l_{\mbox{\tiny ann}}$ takes the form \cite{CarbProt92,CarbProt96}:
\begin{eqnarray*} 
q^2 \sigma^l_{\mbox{\tiny ann}} = (2l+1) 4\pi  
\frac{g_l(\eta) q^{2l+1} C_0^2 (\eta) \mbox{ Im} (- a_l^{cs} ) }{\left| 1 + ig_l(\eta) q^{2l+1} w (\eta) a_l^{cs} \right|^2}.
\end{eqnarray*}
In the limit $q \rightarrow 0$, all partial annihilation cross sections
multiplied by $q^2$ tend to a constant value.

Recall that for $\bar p p$ annihilation the scattering length approximation works up to almost 100 MeV/$c$ antiproton momentum within a few percent accuracy. It also works reasonably well for light nuclei ($^2$H, He). Unfortunately, this is not so for the $\bar p $C system. 
The reason is the large size of the correction from the effective range,  the next term in the expansion~(\ref{ScatLength}). Here we give the following values of scattering length and effective range for S-wave without the Coulomb potential:
\begin{eqnarray*}
\label{equ24}
&&a_0^{\mbox{\tiny s}} =3.22-1.16i \mbox{ fm},\\
&&r_0^{\mbox{\tiny s}} =2.09-0.83i \mbox{ fm}.
\end{eqnarray*}
and with Coulomb potential: 
\begin{eqnarray*}
\label{equ25}
&&a_0^{cs} =-0.24-2.27i \mbox{ fm},\\
&& r_0^{cs} =3.56-3.70i \mbox{ fm}.
\end{eqnarray*}
This approximation works within the same precision for higher partial waves.

\subsection{Choice of parameters in optical potentials}

The choice of parameters in these potential models is not unique. As noted in the literature, there are different ways to choose the depth of complex potential of Woods-Saxon form to fit experimental data. For instance, Ashford {\sl et al.} \cite{Ashford} proposed another choice of the depth of the potential with the imaginary part for C $\sim$ 100 MeV instead of $\sim$ 200 MeV in \cite{FGM2005}. This choice gives practically the same behavior of annihilation cross sections as red curves in Figs. \ref{imgPbarC} and \ref{imgNbarC}. 

Wong {\sl et al.} in 1984 \cite{Wong1984} proposed an even smaller imaginary part of the nuclear potential of $\sim$ 30 MeV. To reproduce the data the authors increased the real part of potential. This kind of potential is shallow in contrast to the initial choice of deep potential corresponding to very strong annihilation. These shallow potentials also described the experimental atomic data and the data on $\bar p A$ annihilation available at that time. Batty {\sl et al.} \cite{BFL1984} argued that the strong annihilation model was favored if one included additional antinucleon data and observed that the $\chi^2$ surfaces in \cite{Wong1984} depend strongly on choice of experimental data included in the fit.

In all these potentials the radius and diffuseness are practically the same and very close to the size of the corresponding nucleus and its experimentally measured diffuseness. 

In 1986 Janouin {\sl et al.}~\cite{Janouin1986} conducted a systematic analysis of a 6 parameter Woods-Saxon optical potential (depth, range and diffuseness for real and imaginary parts of complex potential) and revealed a strong correlation between different fitting parameters needed to describe $\bar p A$ elastic differential cross sections. This phenomenon, known in nuclear physics as the {\sl Igo ambiguity}, was discovered in the optical potential models applied to the $\alpha$-particle scattering~\cite{Igo}. As noted in this study, it is very difficult to find a potential describing at the same time the data from all nuclei at all energies: the authors even proposed a model with diffuseness depending on $A$, $Z$, and $N$ without success.

\begin{figure}[h!] 
\center
\includegraphics [width=14cm]{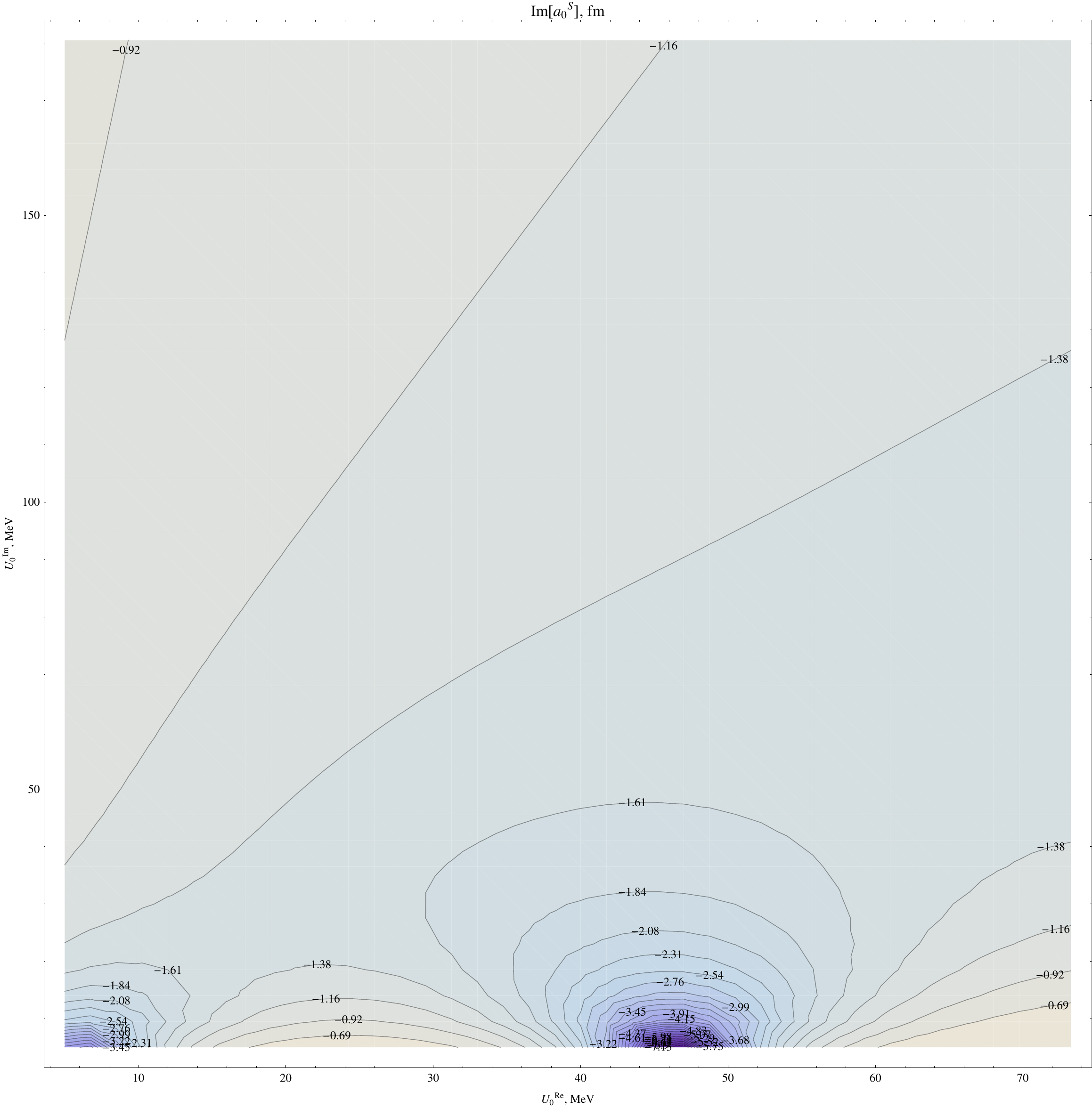}
\caption{Im $a_0^{\mbox{\tiny s}}$ map as a function of the real and imaginary depths of the Woods-Saxon potential.} \label{analyse}
\end{figure}

To illustrate the main difference between models with strong and weak annihilation we present in Fig. \ref{analyse} the imaginary part of the scattering length calculated for a complex Woods-Saxon potential with equal radii and diffusenesses for the real and imaginary parts of potential as a function of two parameters: real and imaginary depths of the potential. The lines on this figure represent constant equal values of Im $a_0^{\mbox{\tiny s}}$. 
For very weak annihilation close to the abscissa axis, one can see a very strong variation of  Im $a_0^{\mbox{\tiny s}}$ due to bound states appearing in this potential with the increase of the real part of the potential, as expected for any attractive real potential.  With increasing annihilation these states (poles of scattering amplitude) are washed out and lose \cite{Shapiro} their simple interpretation as resonance states. Finally, for a deep imaginary potential depth ($\geq100$ MeV), annihilation completely``kills'' the wave function inside nucleus. According to (\ref{strongAnn}), Im $a_0^{\mbox{\tiny s}}$ depends only on the relative phase $\varphi$ of the complex potential. The potential models with strong annihilation are thus quite stable with respect to parameter modifications in contrast to models with weak annihilation.

Recently, Wong and Li proposed \cite{LeeWong2018} a new many parameter potential which describes well the existing experimental data on antinucleon-nucleus annihilation. No comparison with other energy level shifts and widths of atomic levels and differential elastic cross section was presented in the article. To describe the data the authors were forced to choose a very shallow imaginary part for this potential with a quite large value of diffuseness $a_W = 1.050 $ fm, which is practically twice the experimental value \cite{Fricke1995}.  

We will reproduce the Wong and Li good description of the experimental data in a simpler way with fewer parameters and try to understand the consequences of this kind of model for the atomic data and elastic cross section. We propose another choice of parameters in (\ref{equ1}) with a shallower potential, diffuseness scaled by a factor of 2.2, and imaginary part of the potential scaled by a factor of 0.3. The results for this potential with double diffuseness are presented by purple lines in Figs. \ref{imgPbarC} 
and \ref{imgNbarC} show quite impressive agreement with experimental data.

\section{Atomic and elastic scattering data}
Let us discuss other experimental observables.

\subsection{Atomic data and scattering lengths}
There is only one experimental result of $\bar p$C atom measured for D- and F-levels on natural carbon \cite{Roberson77} which gave
\begin{eqnarray*}\label{DFlevels_EXP}
&&\Delta E_{3D}= -4\pm 10 \mbox{eV}, \hspace{1 cm} \Gamma_ {3D}= -42\pm 18 \mbox{eV}; \\
&&\Gamma_{3F}= -0.036^{+0.015}_{-0.011} \mbox{eV}.
\end{eqnarray*}
By use of Trueman formula \cite{Trueman} generalized for higher partial waves in  \cite{Partensky,Lambert,Bakenstoss}, on obtains for D- and F- scattering parameters
\begin{eqnarray}\label{DF_ScattL_EXP}
&&\mbox{Re } a_2^{cs}(\bar p\mbox{C})= 0.5\pm 1.3 \mbox{ fm}^5, \hspace{1 cm} \mbox{Im } a_2^{cs}(\bar p\mbox{C})= -(5.4\pm 2.3) \mbox{ fm}^5; \nonumber \\
&& \mbox{Im } a_3^{cs}(\bar p\mbox{C})= -\left(8.6^{+3.6}_{-2.6} \right) \mbox{ fm}^7.
\end{eqnarray}

In our numerical calculations for strong annihilation and usual diffuseness, these D- and F- scattering parameters $a^{cs}_l$ appear to be equal:
\begin{eqnarray*}
&&a^{cs}_2(\bar p\mbox{C})_{\mbox{\tiny sd}}=1.14-6.39i \mbox{ fm}^5,\\
&&a^{cs}_3(\bar p\mbox{C})_{\mbox{\tiny sd}}=-0.90-3.12i \mbox{ fm}^7.
\end{eqnarray*}
For the potential with double diffuseness, these values are quite different
\begin{eqnarray*}
&&a^{cs}_2(\bar p\mbox{C})_{\mbox{\tiny dd}}= - 33.2 - 57.2 i\mbox{ fm}^5,\\
&&a^{cs}_3(\bar p\mbox{C})_{\mbox{\tiny dd}}= - 72. - 37. i \mbox{ fm}^7.
\end{eqnarray*}
and disagree with atomic data (\ref{DF_ScattL_EXP}). Indices sd and dd refer to simple and double diffuseness of the imaginary part of potentials respectively.

These models give quite different values for the scattering length
\begin{eqnarray*}
a^{cs}_0(\bar p\mbox{C})_{\mbox{\tiny sd}}=-0.24-2.27i \mbox{ fm},\\
a^{cs}_0(\bar p\mbox{C})_{\mbox{\tiny dd}}= - 0.40 - 0.73 i \mbox{ fm}.
\end{eqnarray*}

One observes the same important difference for the scattering length parameters without Coulomb forces corresponding to $\bar n\mbox{C}$ system. The scattering lengths
$a^{s}_{l=0}$ are equal:
\begin{eqnarray*}
a^{s}_0(\bar n\mbox{C})_{\mbox{\tiny sd}}=3.22 - 1.16 i \mbox{ fm},\\
a^{s}_0(\bar n\mbox{C})_{\mbox{\tiny dd}}= 5.36 - 3.57 i\mbox{ fm}.
\end{eqnarray*}
The imaginary part of $a^{s}_0(\bar n\mbox{C})$
is three times larger in the double diffuseness model that for the simpler model. Ashford {\sl et al.}'s model \cite{Ashford} (with weaker but quite strong annihilation and usual diffuseness) gives the imaginary part of $a^{s}_0(\bar n\mbox{C})$ very close to 1 fm:
\begin{eqnarray*}
a^{s}_0(\bar n\mbox{C})_{\mbox{\tiny Ashford}}=3.25 - 1.21 i \mbox{ fm}.
\end{eqnarray*}

\subsection{Elastic scattering}

\begin{figure}[h!] 
\center
\includegraphics [scale=1.3]{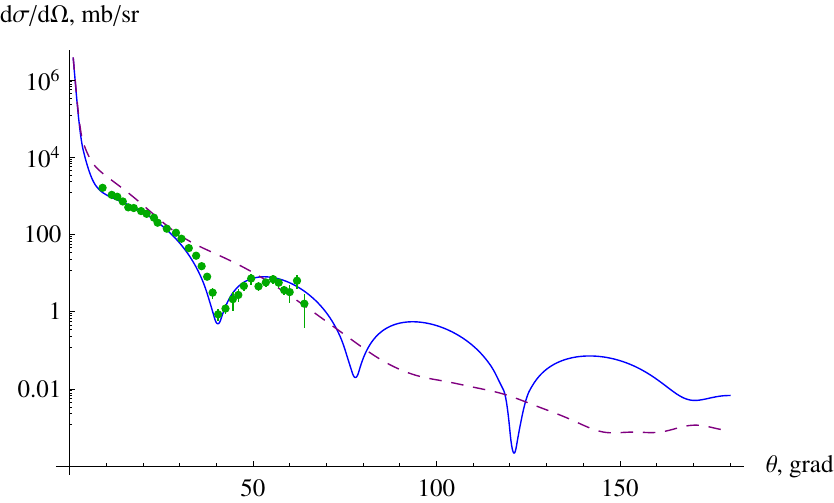}
\caption{$\bar p$C elastic differential cross-section as a function of scattering angle $\theta$ for $p=296$ MeV/c for simple diffuseness (solid line) and double diffuseness potential (dashed line) models. The data is taken from \cite{Garreta1984}.} \label{DiffEl}
\end{figure}

Both approaches with simple and double diffuseness can be compared to the existing data on elastic scattering. The measurement of the elastic differential cross section done by D. Garreta {\sl et al.} extended down to antiproton momentum just below 300 MeV/$c$ ($\bar p$ energy 46.7 MeV) \cite{Garreta1984}. A comparison of the two analyses of the experimental data is presented in Fig. \ref{DiffEl}.
The shallow potential model fails completely to reproduce the experimental data: both differential and total elastic cross sections are wrong. The strength of annihilation is too weak to reduce the elastic cross section to its experimentally observed value. It would be very interesting to measure this observable at lower energies to better constrain parameters of the model.
\begin{figure}[h!] 
\center
\includegraphics [scale=1.3]{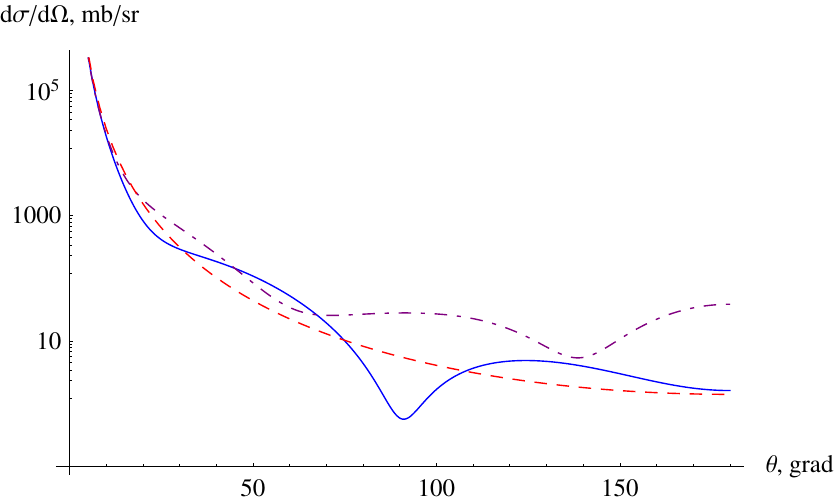}
\caption{Elastic differential cross section 
as a function of scattering angle $\theta$ for $p=100$ MeV/c for simple diffuseness (solid line) and double diffuseness potential (dash-dotted line) models. The pure Coulomb contribution is shown by dashed line.
} \label{imgLowestEn}
\end{figure}

\begin{figure}[h!] 
\center
\includegraphics [scale=1.3]{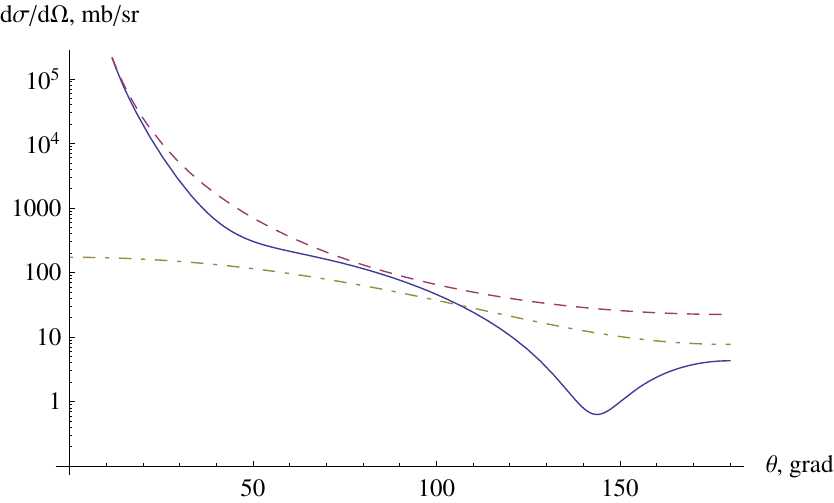}
\caption{Elastic differential cross-section as a function of scattering angle $\theta$ for $p=50$ MeV/c -- solid line; pure Coulomb contribution -- dashed line, pure nuclear contribution calculated for strong annihilation model -- dash-dotted.
} \label{img28}
\end{figure}

As seen from Fig. \ref{imgLowestEn}, the difference between the two models is also important at lower energies, here for $p=100$ MeV/c, and it would be very important to measure the differential cross section at this energy to definitely rule out the double diffuseness model. However at lower energies the contribution from Coulomb scattering becomes more and more important (red dashed line) and the required statistical sensitivity for the differential cross section maybe be difficult to reach. The very strong interference between Coulomb and  nuclear scattering amplitudes may lead to a more sensitive approach.  If the required statistical accuracy in this regime can be reached one can learn about both the real and  imaginary parts of the amplitudes. If at 300 MeV/$c$ the Coulomb amplitude is dominant only in the very forward direction ($\theta \leq 5^\circ$) it starts to hide the nuclear amplitude at lower energies. The relative contributions of these two interactions -- Coulomb (purple) and nuclear (khaki) as well their sum with clear interference effects are presented in Fig.\ref{img28}, where $\bar p A$ elastic differential cross-section in the model of strong annihilation is calculated for $p=50$ MeV/$c$. At 50 MeV/$c$ we are still working with 3 complex amplitudes for the three lowest partial waves. One of them can be determined independently from atomic data.

In the 20--50 MeV/$c$ momentum range, the contributions from two higher partial waves (D and F) represent small corrections to S- and P-partial waves and can be calculated and subtracted from the data by using the scattering length approximation. The F-wave is practically negligible. However, it seems that there is no hope in the near future to measure the nuclear contribution below 20 MeV/$c$. 

To summarize, based on a comprehensive analysis presented above, the predictions of the strong annihilation model with usual
diffuseness is likely to be correct as compared to the model of weak annihilation and double diffuseness. Even if the
strong annihilation model fails to describe some experimental points (not yet confirmed by independent experiments) at
low positive energy it works quite well below threshold (Coulomb states) and for slightly higher energies where one has no doubts on the
precision of experimental data. Moreover, the diffuseness used in this model corresponds perfectly to experimentally observed
one. However, in order to disregard definitely alternative models with double diffuseness, the experimental measurement
of elastic cross section at low energy experiments should be performed.

Nevertheless, as the analysis of the $(n -\bar n)$ oscillation time $\tau_{n \bar n}$ should be carried out in the most conservative manner
possible to completely exclude the possibility of incorrect interpretation of results of such a fundamental experiment, we
will use in our estimations the largest possible value of $\mbox{Im }a_{\bar n A}$, which corresponds to the less realistic model of weak annihilation and double diffuseness potential. We will thus obtain un upper limit on the uncertainty on $\tau_{n \bar n}$.

\section{Consequences for the future ESS experiment}

\subsection{General expression for antineutron flux}

To better understand the impact of uncertainties in our knowledge of the scattering lengths to future neutron oscillation experiments, we start with a general expression for the 
probability of $n-\bar n$ oscillations $P_{n \rightarrow \bar n} (t)$ as a function of the observation time $\tau_{\mbox{\tiny obs}}$ based on the first order Born approximation on the $\varepsilon$ mixing parameter and has the form \cite{NGPSV_Conf} 
\begin{eqnarray}
\label{phenomenology}
P_{n \rightarrow \bar n} (\tau_{\mbox{\tiny obs}}) = \frac{\varepsilon^2}{\omega^2 + \frac{\Gamma_a^2}{4}}\mbox{e}^{-\Gamma_\beta \tau_{\mbox{\tiny obs}}} \left\{1 +\mbox{e}^{-\Gamma_a \tau_{\mbox{\tiny obs}}} - 2\mbox{e}^{-\frac{\Gamma_a}{2} \tau_{\mbox{\tiny obs}}} \cos \omega \tau_{\mbox{\tiny obs}} \right\}.
\end{eqnarray}
where $\Gamma_\beta$ is the inverse neutron $\beta$-decay lifetime (very long with respect to all other characteristic times, so the exponential term will be replaced by 1 in all further discussions), $\Gamma_a$ the $\bar n$ annihilation width, $\varepsilon$ the mixing parameter, and 
$\omega$ a parameter describing the phase shift difference for $n$ and $\bar n$ wave functions accumulated in the $n/\bar n$ guide surface.
$\Gamma_a \equiv \Gamma_a (\vec v ; \mbox{Im }a_{\bar n A}) $ and $\omega \equiv \omega_a (\vec v ; \mbox{Re }a_{\bar n A}) $ depend on $\bar n$ velocity $\vec v$ and are proportional to, respectively, the imaginary and real parts of the scattering length $a_{\bar n A}$ \cite{NGPSV_PRL}.  For instance, the width induced by annihilation of antineutrons on the bottom of the guide, representing the main part of $\bar n$ losses, is related  to Fermi potential of the guide $U_{\bar n} = V_{\bar n} +i W_{\bar n}$ by  
\begin{eqnarray}
\label{width}
\Gamma_a = W_{\bar n} \frac{g\sqrt{\overline{e_{\mbox{\tiny vert}}}}}{\overline{v_{\mbox{\tiny vert}}} V_{\bar n}^{3/2}}.
\end{eqnarray}
Here $\overline{e_{\mbox{\tiny vert}}}$,  $\overline{v_{\mbox{\tiny vert}}}$  are vertical energy and vertical velocity of $\bar n $ averaged over the neutron spectrum,  $g$ is gravitational acceleration. 
The Fermi potential $U_{\bar n} = [(2\pi \hbar^2)/m] (\rho/\mu) b_{\bar n A}$ is directly proportional to $b_{\bar n A}$, with $\rho$ the mass density of material and $\mu$ the atomic mass.

Note that a phase shift difference for $n$ and $\bar n$ wave functions in itself does not destroy the coherence between$n$ and $\bar n$ as evidenced, for instance, by the spectacular preservation of the coherence of polarized neutron spin components in neutron spin-echo spectroscopy \cite{SE}. The point is that, like the neutron-antineutron system, polarized neutron reflection is also a two-state quantum mechanical system and so the optical behavior should be the same. 

The total number $N_{\bar n}$ of $\bar n$ counts in the detector can be evaluated using (\ref{phenomenology}) by integration over all neutron velocities weighted by the distribution function $f(\vec v)$ of neutron speeds used in the experiment 
\begin{eqnarray*}
N_{\bar n} = \int P_{n \rightarrow \bar n} (\tau_{\mbox{\tiny obs}}) f(\vec v) d^3 v.
\end{eqnarray*}

To be able to keep the mirror-reflected neutrons in the quasifree limit $f(\vec v)$ must be shaped so that the maximum transverse velocities  $v_\perp$ are low enough compared to the critical velocities $v_{\mbox{\tiny crit}}$ of the $n/ \bar n$ guide material. This case is discussed in the following subsection. Let us analyze the general expression for the probability of $n - \bar n$ oscillations $P_{n \rightarrow \bar n} (t)$ as a function of time $t$ 
\begin{eqnarray}
\label{phenomenology_anal}
P_{n \rightarrow \bar n} (t) = \frac{\varepsilon^2}{\omega^2 + \frac{\Gamma_a^2}{4}}\mbox{e}^{-\Gamma_\beta t} \left\{1 +\mbox{e}^{-\Gamma_a t} - 2\mbox{e}^{-\frac{\Gamma_a}{2} t} \cos \omega t \right\}.
\end{eqnarray}
We search for maxima of this expression as a function of time $t$ with respect to two parameters $\omega$ and  $\Gamma_a$. We can rewrite (\ref{phenomenology_anal}) as  $P_{n \rightarrow \bar n} = \frac{4\varepsilon^2}{\Gamma_a^2} \mbox{e}^{-\Gamma_\beta t} P(T,R) $ with an auxiliary function
\begin{eqnarray}
\label{optimise}
P(T,R) = \frac{1}{1 +R^2} \left\{1 +\mbox{e}^{-2T} - 2\mbox{e}^{-T} \cos RT \right\},
\end{eqnarray}
where $T= \Gamma_a t/2$ and $R= 2\omega/ \Gamma_a$.
$T$ plays the role of ``effective time''. $R$ is a parameter which we can vary by an appropriate choice of element and its isotopic composition. Let us note that for $R=0$, this function $P(T,0) = 4 \mbox{e}^{-T} \sinh^2 \frac{T}{2}$  has no maximum and is an increasing function of $T$.

The maximum of this function can be found from the condition 
$\frac{\partial }{\partial T }P(T,R) =0$,
which gives the equation to find $T_{\mbox{\tiny max}}$
\begin{eqnarray}
\label{max_eqt}
\cos RT_{\mbox{\tiny max}} + R \sin RT_{\mbox{\tiny max}} = \mbox{e}^{-T_{\mbox{\tiny max}} }.
\end{eqnarray}
The solution to the transcendental equation (\ref{max_eqt}) can be approximated in some limiting cases. Let us start with $R \leq 1$ (or $2\omega \leq \Gamma_a$) and rewrite it in the form
\begin{eqnarray}
\label{max_eqt_new}
\sqrt{1 + R^2} \cos (RT_{\mbox{\tiny max}} -\varphi_R)= \mbox{e}^{-T_{\mbox{\tiny max}} }.
\end{eqnarray}
with $\sin \varphi_R = \frac{R}{\sqrt{1 + R^2}}$. 
For $T_{\mbox{\tiny max}} =0$, the functions in the left and the right side of this equation are equal to 1.
For $T_{\mbox{\tiny max}} > 0$, $\mbox{e}^{-T_{\mbox{\tiny max}} }$ is a rapidly decreasing function whereas $\sqrt{1 + R^2} \cos (RT_{\mbox{\tiny max}} -\varphi_R)$
starts to increases to its maximum value equal to $\sqrt{1 + R^2}$ at $RT_{\mbox{\tiny max}}= \varphi_R$ and then decreasing to zero at $RT_{\mbox{\tiny max}}= \varphi_R+\pi/2$.
Just before becoming zero, this function is equal to the value of the function $\mbox{e}^{-T_{\mbox{\tiny max}} }$.

In this approximation, the first root of eq. (\ref{max_eqt_new}) is 
\begin{eqnarray}
\label{Tmax_1l}
T_{\mbox{\tiny max}}^{(1)} \approx \frac{\pi}{2} \frac{1}{R} + \frac{\varphi_R}{R}.
\end{eqnarray}

This approximation can be improved by expanding this function in the vicinity of this point:
\begin{eqnarray*}
T_{\mbox{\tiny max}}^{(2)} \approx \frac{\pi}{2} \frac{1}{R} + \frac{\varphi_R}{R} - \frac{R}{R\mbox{e}^{\frac{1}{R}\left( \frac{\pi}{2} +\varphi_R\right)}-1}.
\end{eqnarray*}

One can verify that this is a quite small correction. Its maximal value (for $R=1$) is equal to 0.105
to compare to first contributions equal respectively to $\frac{\pi}{2}$ and $\frac{\pi}{4}$. 

Let us emphasize that, for usual time variable $t$, this maximum
corresponds to (\ref{Tmax_1l})
$t_{\mbox{\tiny max}} ^{(1)} \approx \frac{\pi}{2} \frac{1}{\omega} + \frac{2}{\Gamma_a}$
or 
\begin{eqnarray}
\label{tau_optimum}
\tau_{\mbox{\tiny obs}}^{\mbox{\tiny max}} \approx \frac{1}{4} \tau_{\omega} + 2 \tau_{\bar n},
\end{eqnarray}
where we introduced $\tau_{\omega} = 2\pi / \omega$. 
\begin{figure}[h!] 
\center
\includegraphics [width=10cm]{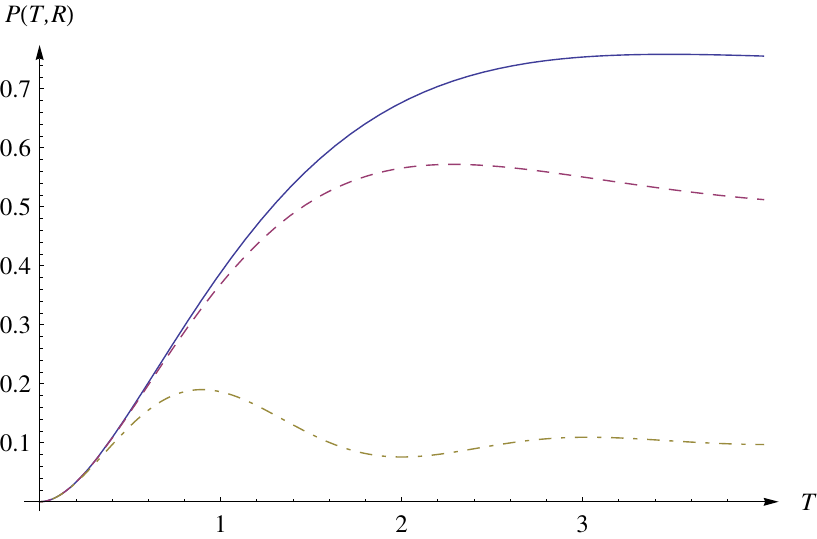}
\caption{Probability of $n \bar n$ oscillations (\ref{optimise}) for $R=0.6$ (solid line), $R=1$ (dashed line), and $R=3 $ (dash-dotted line) as a function of time.} \label{t06}
\end{figure}

Equation (\ref{max_eqt}) for $R > 1$ (or $2\omega >\Gamma_a$) can be solved in an analogous way, by introducing new variables
$Q = 1/R= \Gamma_a/2\omega$ and $\tilde T = T/Q = \omega t$.
By using the same conventions, and with the same precision as in (\ref{Tmax_1l}), one obtains
\begin{eqnarray*}
\tilde T_{\mbox{\tiny max}}^{(1)} \approx \frac{\pi}{2} +\arcsin \frac{1}{\sqrt{1 + Q^2}}.
\end{eqnarray*}

In the limit $Q \rightarrow 0$,  $ \tilde T_{\mbox{\tiny max}}^{(1)} \rightarrow \pi$,
as expected from the initial equation (\ref{phenomenology}) for zero annihilation width
\begin{eqnarray}
\label{phenomenology_without_ann}
P_{n \rightarrow \bar n} (t) = \frac{4 \varepsilon^2}{\omega^2}\mbox{e}^{-\Gamma_\beta t} \sin^2 \frac{T}{2}.
\end{eqnarray}

\begin{figure}[h!] 
\center
\includegraphics [width=10cm]{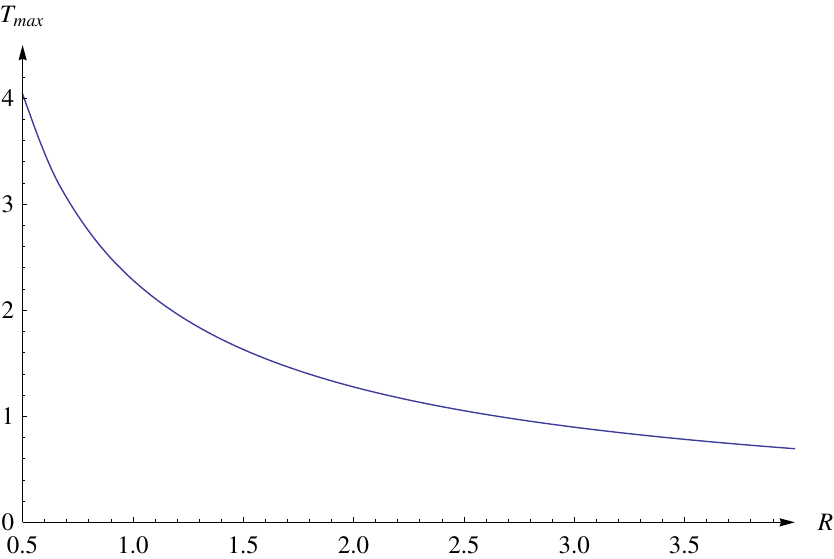}
\caption{$T_{\mbox{\tiny max}}$ as a function of $R$.} \label{final}
\end{figure}

The results of numerical solution of eq.(\ref{max_eqt}) are presented in fig. \ref{final}.

These general expressions allow one to optimize the guide length and the choice of materials for future
experiments.

\subsection{ESS experiment}
In the proposed ESS experiment, the $n/\bar n$ guide length is short enough that the observation time $\tau_{\mbox{\tiny obs}} \sim 0.4$ s is
smaller than $\tau_{\bar n} =1/\Gamma_a \sim $2 s and $\tau_{\omega} =2\pi/\omega \sim $ a few seconds for any neutron entering
the guide and we are far from this optimal value $\tau_{\mbox{\tiny obs}}
\ll \tau_{\mbox{\tiny obs}}^{\mbox{\tiny max}}$ from (\ref{tau_optimum}).

In this limit, the expression (\ref{phenomenology}) reduces to
\begin{eqnarray}
\label{phen_short}
P_{n \rightarrow \bar n} (\tau_{\mbox{\tiny obs}}) = \left( \frac{\tau_{\mbox{\tiny obs}}}{\tau_{n \bar n}} \right)^2
\left(1 - \frac{\Gamma_a}{2} \tau_{\mbox{\tiny obs}} +...\right),
\end{eqnarray}
where we include only the first order correction in $t$. The term with $\omega$ will appears in the next order.

This expression allows us to determine the relative uncertainty $\Delta \tau_{n \bar n}/ \tau_{n \bar n}$ which, in this approximation, is related
to the relative uncertainty on the imaginary part of the scattering length $\Delta \mbox{Im }a_{\bar n A}/ {\mbox{Im }a_{\bar n A}}$ by 
\begin{eqnarray}
\label{Uncertainty}
\frac{\Delta \tau_{n \bar n}} {\tau_{n \bar n} } = \frac{1}{4} \frac{\Delta \mbox{Im }a_{\bar n A} }{\mbox{Im }a_{\bar n A}}\frac{\tau_{\mbox{\tiny obs}} }{\tau_{\bar n}}. 
\end{eqnarray}
Here, we used the fact that $\Gamma_a$ is proportional to $\mbox{Im } a_{\bar n A}$ as it can be seen from eq.(\ref{width}). Note once more that the uncertainty on the real part will appear in the next order of the expansion.

As we noted in the introduction, the most critical parameter for designing a $n-\bar n$ oscillation experiment using guide reflection is the uncertainty of the knowledge of the $a_{\bar n A}$. To minimize the associated uncertainties, one can use the fact that the predictive power of the present theory is higher for the light nuclei than for heavy nuclei. Deuterium is a good material for neutron reflection, and $a_{\bar n ^2\mbox{\tiny H}}$ is well known. There are many $^2$H-rich compounds that can be used to make $n/\bar n$ guides. Since the uncertainty in $a_{\bar n A}$ increases with increasing $A$, one should look for technologically convenient materials containing a maximum fraction of $^2$H atoms plus light elements (carbon, oxygen etc).

Another option is carbon. With the $\sim 10 \%$ uncertainty in $ \mbox{Im }a_{\bar n \mbox{\tiny C}}$ provided by the model of strong $\bar n$ annihilation, the corresponding systematic effect on the oscillation probability is only $\sim 0.5\%$ and can be neglected. Within this model and for the parameters of the $n$ beam and the experiment length at ESS, one can assume with a good accuracy that (nearly) all $\bar n$ reach the annihilation detector. 

Until the  alternative double diffuseness potential model is completely eliminated in the low-energy limit by new experimental data as we expect based on our analysis, we have to use the largest uncertainty. This means that the lifetime of antineutrons $\tau_{\bar n}$ in the $n /\bar n$ guide can be as much as three times smaller. In this case a large fraction of $\bar n$ can be lost in the guide on the way to the annihilation detector, and a systematic uncertainty in the estimation of $\tau_{n \bar n}$ increases to  $\sim 15\%$ using eq.(\ref{Uncertainty}) and for the mean values for the neutron velocity, the maximum possible length of the experiment, and sufficiently low normal components of velocities so that energy-dependent losses are negligible. This estimation can easily be done precisely given $f(\vec v)$. The resulting uncertainty in the estimation of $\tau_{n \bar n}$ is at the borderline of acceptability. The higher density of carbon in a diamond-like coating would make the real part of the scattering length is proportionally larger than that for the carbon parameters assumed above. 

Another option for guide material is deuterated polyethylene or any technologically convenient material which contains a significant fraction of deuterium atoms plus some light elements. In the first approximation, the scattering length for $\bar n- ^2$H interaction can be extracted nearly directly from experiment. The fact that the scattering length for 2 out of 3 atoms is known quite well, means that the overall uncertainty for the estimation of $\tau_{\bar n}$ in a deuterated-polyethylene $n /\bar n$ guide is 3 times smaller. In this case the overall uncertainty of estimating $\tau_{n \bar n}$ is roughly 5$\%$. all else being equal. This value can be treated as a simple systematic uncertainty along with the other systematic uncertainties in this experiment. 

We can conclude that at least one type of material can be used for the construction of a $n/\bar n$ guide for the $n - \bar n$ oscillation experiment at ESS given the current status of theory and experiment for antineutron-nucleus scattering lengths. Neutron guides made of the materials discussed above have been used successfully in numerous experiments with long storage of ultracold neutrons on specular trajectories \cite{Arzumanov2011,Nesvizhevsky2007}. These technologies can be extended to cold and very cold neutron guides, which do not require so many consequent reflections so that the parameters can be even more relaxed. 

\section{Conclusions}

We analyzed different theoretical approaches to extract the imaginary part of the antineutron-nucleus scattering length. For deuterium the scattering length can be extracted almost directly from the experimental data. For heavier nuclei, one has to use some potential models. Among the existing potentials describing experimental data on the antinucleon-nucleus interaction, we conclude that the model with strong annihilation and simple diffuseness is favored despite its disagreement with some of the scarce existing experimental data: one measurement for $\bar p$C annihilation and a small number of measurements for different nuclei obtained by one experimental group. It would be very interesting to obtain predictions for antineutron-nucleus scattering length in some alternative theoretical approaches, for instance, based on chiral effective field theory proposed recently \cite{Dai}.

Further experimental study of the antinucleon-nucleus interaction at lowest possible energies would be of vital importance to understand the exiting discrepancies between some experimental data and the model of strong annihilation and to better constrain the imaginary part of the antineutron-nucleus scattering length. In particular, it would be very important to measure $\bar p$C differential elastic cross section at low energies (for 5 MeV $\bar p$) at the Antiproton Decelerator at CERN. A special theoretical and experimental effort to determine better the real part of the $\bar n^2$H scattering length would be also very useful. 

Within the strong annihilation model, the imaginary part of antineutron-nucleus scattering length is practically the same for all nuclei and is equal to $\sim$ 1 fm. Under the most conservative assumption using a less realistic model which describes annihilation data but is not able to describe other experimental data, this upper limit increases to 3 fm. With this value of the imaginary part of the antineutron-nucleus scattering length, one can calculate the most conservative constrain for the sensitivity of the $n -\bar n$ oscillation experiment. 

The relative uncertainty in evaluation of the $n-\bar n$ oscillation time is a factor of $\sim$ 20 smaller than the relative uncertainty in the value of the imaginary part of the antineutron-nucleus scattering length in the quasifree limit and for the materials we considered for $n/\bar n$ guide (see eq.(\ref{Uncertainty}) using parameters relevant to the proposed ESS $n -\bar n$ oscillation experiment). This insensitivity to the precise value of the antineutron-nucleus scattering length shows that one can perform realistic design of the future experiment despite the lack of direct measurements of this quantity.

We considered three options for the material of $n$ and $\bar n$ guide. For the carbon coating, the most conservative estimation for the systematic uncertainty in the evaluation of the $n -\bar n$ oscillation time caused by the uncertainty of knowledge of the $\bar n$C scattering length is $\sim~15\%$, slightly larger when energy-dependent corrections are taken into account. A diamond-like carbon coating has a larger real part of 
the Fermi potential compared to natural carbon which reduces this uncertainty. The uncertainty in the estimation of the $n -\bar n$ oscillation time for guide material dominated by $^2$H with all else being equal falls to $\sim 5\%$ due to the better knowledge of the $\bar n^2$H scattering length. Such an uncertainty might be acceptable for designing such an experiment. As all these materials have been used successfully for ultracold $n$ guides, there are good prospects of using them also for very cold and cold neutrons. Many other options for the neutron guide material can be considered provided a better knowledge of the antineutron-nucleus scattering length is achieved for this material.

It would also be valuable to conduct a neutron optical analysis in the other extreme case in which one assumes that the antineutron scattering lengths are known for all stable nuclei to some specified precision and take the range of values predicted by the various potential models. Such an analysis could help define the envelope of the most extreme possibilities for this mode of conduct for a free neutron-antineutron oscillation experiment.  

In the future, we plan to analyze quantitatively various options for designing the proposed neutron-antineutron oscillation experiment at ESS, which
takes into account neutron extraction from the spallation source (with or without focusing by supermirrors \cite{Mezei1977} or/and nanodiamond reflectors \cite{Nesvizhevsky2018}), different neutron spectra provided by a source of cold or very cold neutrons, and different guide geometries.

\section{Acknowlegments}

We are grateful to Albert Young and Ken Andersen for pointing the possibility to use diamond-like coating for the $n/ \bar n$ guide in the  $n - \bar n$ oscillation experiment at ESS, as well as to Luca Venturelli, David Milstead, and Gustaaf Brooijmans for discussions of prospects of better experimental constraints for antineutron-nucleus scattering length. W. M. Snow acknowledges support from US National Science Foundation grants PHY-1614545, PHY-1913789 and by the Indiana
University Center for Spacetime Symmetries. 
V. Gudkov acknowledges support from the U.S. Department of Energy Office of Science, Office of Nuclear Physics program under Award No. DE-SC0020687.
The authors are grateful to the participants of INT-17-69W Workshop ``Neutron-antineutron oscillations: appearance, disappearance and baryogenesis" in Seattle, USA, held on 23-27 October 2017.

\bibliography{PbarNucleus}

\begin{thebibliography}{99}

\bibitem{Stueckelberg38}
E.C.G. Stueckelberg, Helv. Phys. Acta \textbf{11}, 312 (1938).

\bibitem{Sakharov67}
A.D. Sakharov, JETP. Lett. \textbf{5}, 24 (1967).

\bibitem{Kuzmin70}
V.A. Kuzmin, JETP. Lett. \textbf{13}, 335 (1970).

\bibitem{Georgi74}
H. Georgi et al., Phys. Rev. Lett. \textbf{32}, 438 (1974).

\bibitem{Zeldovich76}
Ya.B. Zeldovich, Phys. Lett. A \textbf{59}, 254 (1976).

\bibitem{Hooft78}
G.t'Hooft, Phys. Rev. D \textbf{14}, 3432 (1976).

\bibitem{Davidson1979}
A. Davidson, Phys. Rev. D \textbf{20}, 776(1979).

\bibitem{MohapatraPRL80}
R.N. Mohapatra and R. E. Marshak, Phys. Rev. Lett. \textbf{44}, 1316 (1980).

\bibitem{Kazarnovskii80}
M.V. Kazarnovskii, V.A. Kuz'min, K.G. Chetyrkin, M.E. Shaposhnikov,  JETP Lett. \textbf{32}, 82 (1980).

\bibitem{Kuo80}
T.K. Kuo and S. T. Love, Phys. Rev. Lett. \textbf{45}, 93 (1980).

\bibitem{Chang80}
L.N. Chang and N.P. Chang, Phys. Lett. B \textbf{92}, 103 (1980).

\bibitem{MohapatraPLB80}
R.N. Mohapatra and R. E. Marshak, Phys. Lett. B \textbf{94}, 183 (1980).

\bibitem{Chetyrkin81}
K.G. Chetyrkin, M.V. Kazarnovsky, V.A. Kuzmin, M.E. Shaposhnikov, Phys. Lett. B \textbf{99}, 358 (1981).

\bibitem{Dolgov81}
A. D. Dolgov and Ya. B. Zeldovich, Rev. Mod. Phys. \textbf{53}, 1 (1981).

\bibitem{Cowsik81}
R. Cowsik and S.Nussinov,  Phys. Lett. B \textbf{101}, 237 (1981).

\bibitem{Rao82}
S. Rao and R.Shrock, Phys. Lett. B \textbf{116}, 238 (1982).

\bibitem{Misra83}
S.P. Misra and U. Sarkar, Phys. Rev. D \textbf{28}, 249 (1983).

\bibitem{Rao84}
S. Rao and R.E.Shrock, Nucl. Phys. B \textbf{232}, 143 (1984).

\bibitem{Kuzmin85}
V.A. Kuzmin, V.A.Rubakov, M.E.Shaposhnikov, Phys. Lett. B \textbf{155}, 36 (1985).

\bibitem{Fukugita86}
M. Fukugita  and T.Yanagida,  Phys. Lett. B \textbf{174}, 45 (1986).

\bibitem{Shaposhnikov87}
M.E. Shaposhnikov, Nucl. Phys. B \textbf{287}, 757 (1987).

\bibitem{Dolgov92}
A.D. Dolgov, Phys. Rep. \textbf{222}, 309 (1992).

\bibitem{Huber01}
S.J. Huber and Q.Shafi, Phys. Lett. B \textbf{512}, 365 (2001).

\bibitem{Babu01}
K.S. Babu and R.N Mohapatra, Phys. Lett. B \textbf{518}, 269 (2001).

\bibitem{Nussinov02}
S. Nussinov and R. Shrock, Phys. Rev. Lett. \textbf{88}, 171601 (2002).

\bibitem{Babu06}
K.S. Babu, R.N. Mohapatra, S. Nasri, Phys. Rev. Lett. \textbf{97}, 131301 (2006).

\bibitem{Dutta06}
B. Dutta, Y. Mimura, and R. N. Mohapatra, Phys. Rev. Lett. \textbf{96}, 061801 (2006)

\bibitem{Berezhiani06}
Z. Berezhiani and L. Bento, Phys. Rev. Lett. \textbf{96}, 081801 (2006).

\bibitem{Bambi07}
C. Bambi, A.D.Dolgov, K.Freese, Nucl. Phys. B \textbf{763}, 91 (2007).

\bibitem{Babu09}
K. S. Babu, P. S. Bhupal Dev, and R. N. Mohapatra, Phys. Rev. D \textbf{79}, 015017 (2009).

\bibitem{Mohapatra09}
R.N. Mohapatra, J. Phys. G \textbf{36}, 104006 (2009).

\bibitem{Dolgov10}
A.D. Dolgov, Phys. At. Nucl. \textbf{73}, 558 (2010).

\bibitem{Gu11}
P.-H. Gu, Phys. Lett. B \textbf{705}, 170 (2011).

\bibitem{Morrissey12}
D.E. Morrissey and M.J. Ramsey-Musolf, New J. Phys. \textbf{14}, 125003 (2012).

\bibitem{Arnold13}
 J.M. Arnold, B. Fornal, and M.B. Wise, Phys. Rev. D \textbf{87}, 075004 (2013).

\bibitem{Babu13}
K. S. Babu, P. S. Bhupal Dev, Elaine C. F. S. Fortes, and R. N. Mohapatra, Phys. Rev. D \textbf{87}, 115019 (2013).

\bibitem{Canetti13}
L. Canetti, M. Drewes, T. Frossard, and M. Shaposhnikov, Phys. Rev. D \textbf{87}, 093006 (2013).

\bibitem{Gouvea14}
A. de Gouvêa, J. Herrero-García, and A. Kobach, Phys. Rev. D \textbf{90}, 016011 (2014).

\bibitem{Berezhiani16}
Z. Berezhiani, Europ. J. Phys. C \textbf{76}, 705 (2016).

\bibitem{Grojean18}
C. Grojean, B.Shakya, J.D. Wells, and Z. Zhang, Phys. Rev. Lett. \textbf{121}, 171801  (2018).

\bibitem{Berezhiani2019}
Z. Berezhiani, R. Biondi, Yu. Kamyshkov, and L. Varriano, Physics \textbf{1}, 271 (2019).

\bibitem{Shrock2019}
S. Girmohanta  and R. Shrock, Phys. Rev. D \textbf{100}, 115025 (2019).

\bibitem{Shrock_2_2019}
S. Girmohanta  and R. Shrock, Phys. Lett. \textbf{B803}, 135296 (2020).

\bibitem{Shrock_3_2019}
S. Girmohanta  and R. Shrock, Phys. Rev. D \textbf{101}, 015017 (2020).

\bibitem{Dover83}
C.B. Dover, A. Gal, and J.M. Richard, Phys. Rev. D \textbf{27}, 1090 (1983).

\bibitem{Dover89}
C.B. Dover, A. Gal, and J.M. Richard Nucl. Instr. Meth. A \textbf{284}, 13 (1989).

\bibitem{Gal00}
A. Gal, Phys. Rev. C \textbf{61}, 028201 (2000).

\bibitem{Chung02}
J. Chung, W. W. M. Allison, G. J. Alner, D. S. Ayres, W. L. Barrett, P. M. Border, et al., Phys. Rev. D \textbf{66}, 032004 (2002).

\bibitem{Friedman08}
E. Friedman and A. Gal, Phys. Rev. D \textbf{78}, 16002 (2008).

\bibitem{Abe15}
K. Abe et al., Phys. Rev. D \textbf{91}, 072006 (2015).

\bibitem{Baldo94}
M. Baldo-Ceolin et al., Zeit. Phys. C \textbf{63}, 409 (1994).

\bibitem{NGPSV_Conf} V.V. Nesvizhevsky, V. Gudkov, K.V. Protasov, W.M. Snow and A.Yu. Voronin, EPJ WEb of Conferences {\bf 191}, 01005 (2018).

\bibitem{NGPSV_PRL} V.V. Nesvizhevsky, V. Gudkov, K.V. Protasov, W.M. Snow and A.Yu. Voronin, Phys. Rev. Lett. {\bf 122}, 221802 (2019).

\bibitem{NGPSV_PLB} V.V. Nesvizhevsky, V. Gudkov, K.V. Protasov, W.M. Snow and A.Yu. Voronin, Phys. Lett. {\bf B803}, 135357 (2020) .

\bibitem{Addazi2020} A. Addazi {\it et al.}, arXiv: 2006.04907 / physics.ins-det (2020).

\bibitem{Gudkov2020} V. Gudkov, et. al., Phys. Lett. {\bf B808}, 135636 (2020).

\bibitem{Augsburge1999} M. Augsburger et al., Phys. Lett. {\bf B461}, 417 (1999).

\bibitem{Zenoni1999} A. Zenoni et al., Phys. Lett. {\bf B461}, 413 (1999).

\bibitem{Protasov2000} K.V. Protasov, G. Bonomi, E. Lodi Rizzini, A. Zenoni, Eur. Phys. J. {\bf A 7}, 429 (2000).

\bibitem{Popov81} V.S. Popov, A.E. Kudryavtsev, V.I. Lisin, V.D. Mur, Sov. Phys. JETP, {\bf 53}, 850 (1981).

\bibitem{Mur83} V.D. Mur, A.E. Kudryavtsev, V.S. Popov, Sov. J. Nucl. Phys. {\bf 37}, 844 (1983).

\bibitem{CarbProt92} J. Carbonell, K. Protasov, J. Phys. G: Nucl. Part. Phys., {\bf 18}, 1863 (1992).

\bibitem{WGN1985}  S. Wycech, A.M. Green, and J.A. Niskanen, Phys. Lett, {\bf B152}, 308 (1985).

\bibitem{LT1990}  G.P. Latta, P.C. Tandy, Phys. Rev. C  {\bf 42}, R1207 (1990). 

\bibitem{LRW1991}   G.Q. Liu, J.M. Richard, S. Wycech, Phys. Lett. {\bf  B 260} 15 (1991) .

\bibitem{Yan2008}  Y. Yan et al., Phys. Lett, {\bf B659}, 555 (2008).

\bibitem{KW} M. Kohno, W. Weise, Nucl. Phys. {\bf A 454}, 429 (1986); Nucl. Phys. {\bf A 479}, 433c (1989).

\bibitem{Batty89} C.J. Batty, Rep. Prog. Phys. {\bf 52}, 1165 (1989).

\bibitem{Batty95} C.J. Batty, E. Friedman, A. Gal, Nucl. Phys., {\bf A592}, 487 (1995).

\bibitem{Batty01} C.J. Batty, E. Friedman, A. Gal, Nucl. Phys., {\bf A689}, 721 (2001).

\bibitem{FGM2005} E. Friedman, A. Gal, J. Mares, Nucl. Phys., {\bf A761}, 283 ( 2005).

\bibitem{Fricke1995} G. Fricke, C. Bernhardt, K. Heilig, L. A. Schaller, L. Schellenberg, E. B. Shera, C. W. de Jager, at. {\it Data Nucl. Data Tables} {\bf 60}, 177 (1995).

\bibitem{Friedman1999} E. Friedman and A. Gal, Nucl. Phys. {\bf A658}, 345 (1999).

\bibitem{KPV2000} V.A. Karmanov, K.V. Protasov, and A.Yu. Voronin, Eur. Phys. J. {\bf A 8}, 429 (2000).

\bibitem{Bressani2003} T. Bressani and A. Filippi, Phys. Rep. {\bf 383}, 213 (2003).


\bibitem{Friedman2014} E. Friedman, Nucl. Phys. {\bf A 925}, 141 (2014).

\bibitem{Friedman2015} E. Friedman, Hyperfine Interact {\bf 234}, 77 (2015).

\bibitem{Aghai2018} H. Aghai-Khozani et al., Nucl. Phys. {\bf A 970}, 366 (2018).

\bibitem{Nakamura1984} K. Nakamura, et al., Phys. Rev. Lett. {\bf 52}, 731 (1984).

\bibitem{Astrua2002} M. Astrua et al., Nucl. Phys. {\bf A 697}, 209 (2002).

\bibitem{Mur85} V.D. Mur, V.S. Popov, Sov. J. Nucl. Phys. {\bf 42}, 930 (1985).

\bibitem{CarbProt96} J. Carbonell, K. Protasov, Z Phys. {\bf A355}, 87 (1996).

\bibitem{Ashford} V. Ashford, M. E. Sainio, M. Sakitt, J. Skelly, R. Debbe, W. Fickinger, R. Marino, D. K. Robinson, Phys. Rev. {\bf C30}, 1080  (1984);
Phys. Rev. {\bf C31}, 663 (1985).

\bibitem{Wong1984} C.-Y. Wong, A. K. Kerman, G. R. Satchler, A. D. MacKellar, Phys. Rev. {\bf C29}, 574 (1984).

\bibitem{BFL1984} C.J. Batty, E. Friedman, J. Lichtenstadt, 
Phys. Lett., {\bf A142B}, 241 (1984);
Nucl. Phys., {\bf A436}, 621 (1985).

\bibitem{Janouin1986} S. Janouin et al., Nucl. Phys. {\bf A 451}, 541 (1986).

\bibitem{Igo} G. Igo, Phys. Rev. Lett. {\bf 1} (1958) 72, Phys. Rev. {\bf 115}, 1665 (1950).

\bibitem{Shapiro} J. Carbonell, O.D. Dalkarov, K.V. Protasov, I.S. Shapiro, Nucl. Phys., {\bf A535}, 651 (1991).

\bibitem{LeeWong2018} T.-G. Lee, C.-Y. Wong, Phys. Rev. {\bf C97}, 054617 (2018).

\bibitem{Roberson77} P. Roberson et al., Phys. Rev. {\bf 16}, 1945 (1977).

\bibitem{Trueman} T.L. Trueman, Nucl. Phys., {\bf 26}, 57 (1961).

\bibitem{Partensky} A. Partensky and M. Ericson, Nucl. Phys., {\bf B1}, 382 (1967).

\bibitem{Lambert} E. Lambert, Helv. Phys. Acta, {\bf 42}, 667 (1969).

\bibitem{Bakenstoss} G. Bakenstoss, Ann. Rev. Nucl. Sc. {\bf 20}, 467 (1970).

\bibitem{Garreta1984} D. Garreta et al., Phys. Lett. {\bf B 135}, 266 (2011).

\bibitem{SE} F. Mezei, Z. Phys. {\bf A 255}, 146 (1972). 

\bibitem{Arzumanov2011} A.S. Arzumanov, L.N. Bondarenko, P. Geltenbort, V.I. Morozov, V.V. Nesvizhevsky, Yu.N. Panin, A.N. Strepetov, D.Yu. Chuvilin, Crystall. Rep. {\bf 56}, 1197 (2011).

\bibitem{Nesvizhevsky2007} V.V. Nesvizhevsky, G. Pignol, K.V. Protasov, G. Quemener, D. Forest, P. Ganau, J.M. Mackowski, Ch. Michel, J.L. Montorio, N. Morgado, L. Pinard, A. Remillieux, Nucl. Instr. Meth. {\bf A 578}, 435 (2007).

\bibitem{Dai} L.-Y. Dai,  J. Haidenbauer,  U.-G. Meissner   JHEP  {\bf 07}, 078  (2017).

\bibitem{Mezei1977}  F. Mezei, P.A. Dagleish, Comm. Phys.  {\bf  2}, 41 (1977).

\bibitem{Nesvizhevsky2018} V.V. Nesvizhevsky, U. Koster, M. Dubois, N. Batisse, L. Frezet, A. Bosak, L. Gines, O. Williams,  Carbon  {\bf 130}, 799 (2018).

\end{thebibliography}

\end{document}